\documentclass[conference,a4paper]{IEEEtran}
\usepackage[left=1.5cm,right=1.5cm,top=2cm,bottom=2cm,a4paper]{geometry}
\usepackage{amsmath}
\usepackage{array}
\usepackage{amsthm}
\usepackage{bbm}
\usepackage{lipsum}

\newtheorem{thm}{Theorem}
\newtheorem{lem}{Lemma}
\newtheorem*{thm*}{Theorem}

\newcommand\blfootnote[1]{%
  \begingroup
  \renewcommand\thefootnote{}\footnote{#1}%
  \addtocounter{footnote}{-1}%
  \endgroup
}
\ifCLASSINFOpdf
\else
\fi
\usepackage{amsmath}
\usepackage{array}
\begin{document}
%
\title{Community Detection with Colored Edges}

\author{\IEEEauthorblockN{Narae Ryu}
\IEEEauthorblockA{School of EE, KAIST, Daejeon, Korea\\
Email: nrryu@kaist.ac.kr}
\and
\IEEEauthorblockN{Sae-Young Chung}
\IEEEauthorblockA{School of EE, KAIST, Daejeon, Korea\\
Email: schung@kaist.ac.kr}}


%

\onecolumn
\pagenumbering{arabic}

\maketitle
\blfootnote{The material in this paper was presented in part at the IEEE International Symposium on Information Theory (ISIT) 2016.}
\begin{abstract}
In this paper, we prove a sharp limit on the community detection problem with colored edges. We assume two equal-sized communities and there are $m$ different types of edges. If two vertices are in the same community, the distribution of edges follows $p_i=\alpha_i\log{n}/n$ for $1\leq i \leq m$, otherwise the distribution of edges is $q_i=\beta_i\log{n}/n$ for $1\leq i \leq m$, where $\alpha_i$ and $\beta_i$ are positive constants and $n$ is the total number of vertices. Under these assumptions, a fundamental limit on community detection is characterized using the Hellinger distance between the two distributions. If $\sum_{i=1}^{m} {(\sqrt{\alpha_{i}}-\sqrt{\beta_{i}})}^{2}>2$, then the community detection via maximum likelihood (ML) estimator is possible with high probability. If $\sum_{i=1}^{m} {(\sqrt{\alpha_{i}}-\sqrt{\beta_{i}})}^{2}<2$, the probability that the ML estimator fails to detect the communities does not go to zero. 
\end{abstract}


%
\IEEEpeerreviewmaketitle

\section{Introduction}

In community detection, the community structure is detected from a given graph by observing the relations between the vertices. Recently, the community detection problem is getting popular in many research fields, such as biology, computer science, and sociology \cite{Fortunato:2010}.

The limit on the community detection is proven in many cases. For instance, for the case of $2$ communities and general $k$ communities on the stochastic block model (SBM), the limit is known when the probability is of order of $\log{n}/n$ \cite{Abbe:2016}, \cite{Abbe:2015}. Even when the parameters of SBM are unknown, the limit is proven in \cite{AS:2015}. Also, there are works about recovering a hidden community. In \cite{Hajek:2015}, they provide nearly matching necessary and sufficient conditions for the recovery of densely subgraph when the distribution of edges follows Bernoulli and Gaussian. Our main theorems generalize the corresponding results in \cite{Abbe:2016}. Also, \cite{Jog:2015} proved the same results as this paper. However, our proofs are based on Cramer's theorem and very simple.
	
We consider a graph $G=(V,E)$ where $V$ is the set of $n$ vertices and $E$ is the set of edges that connects the vertices. Here, we focus on the case where the graph has two communities of equal size. Unlike most papers on SBM, we will deal with a more general version of SBM that contains colored edges. In other words, two vertices are connected with edge with color $1,2,...,m$. From now on, an edge with color $i$ is denoted by $i$-edge for simplicity. Therefore, the probability that two vertices within the same community are connected with an $i$-edge is $p_i$ and they are disconnected, i.e., no edge, with probability $1-\sum_{i=1}^{m}p_i$. In a similar way, two vertices in different communities are connected with an $i$-edge with probability $q_i$ and disconnected with probability $1-\sum_{i=1}^{m}q_i$.
	
Furthermore, we assume that $p_i$ and $q_i$ are of order of $\log{n}/n$. Hence, we set $p_i$ and $q_i$ as $p_i=\alpha_i\log{n}/n$ and $q_i=\beta_i\log{n}/n$ where $\alpha_i$ and $\beta_i$ are positive constants. The reason why we choose such a probability is to guarantee the connectivity of the graph. According to \cite{Erdos:1960}, if $\sum_i \alpha_i + \sum_i \beta_i<2$, then there would be an isolated vertex with high probability.
	
In this paper, we prove a fundamental limit on the SBM with colored edges. As the maximum likelihood (ML) detector is optimal in a sense of minimizing the probability of error, we first specify what the ML rule is in this model. In chapter 3, we provide the limit on the detection of community via two theorems. Theorem 1 proves the sufficient condition for the detection. Theorem 2 proves the necessary part. Therefore, by combining these two theorems, we can get the sharp limit on the community detection with colored edges.

\section[Maximum Likelihood Estimator]{Maximum Likelihood Estimator on SBM}
As mentioned before, we have 2 communities of size $n/2$ each. If all edges have the same color and $p>q$, then the ML rule is simply to find the community that has the minimum number of edges across two communities. However, if we have $m$ different types of edges, the rule becomes slightly different. Suppose we fix a graph $G$ and the number of $j$-edges inside the graph $G$ is denoted by $k_j$. Then, we need to find the communities $A$ and $B$ of equal size, which maximizes $\Pr(G|A,B)$ among the $2^n$ possible community assignments. To calculate $\Pr(G|A,B)$, let's fix some communities $A$ and $B$ and define $l_j$ as the number of inner $j$-edges inside the communities $A$ and $B$. Then, the number of $j$-edges across the communities is $k_j-l_j$, obviously. Therefore, $\Pr(G|A,B)$ is,

\begin{eqnarray*}
\Pr(G|A,B) &=& p_1^{l_1}p_2^{l_2} \cdots p_m^{l_m}\left(1-\sum_{i=1}^{m} p_i\right)^{2 {n/2 \choose 2}-\sum_{i=1}^m l_i} q_1^{k_1-l_1} q_2^{k_2-l_2} \cdots q_m^{k_m-l_m} \left(1-\sum_{i=1}^{m} q_i\right)^{\frac{n^2}{4}-\sum_{i=1}^m (k_i-l_i)} \\
&=& \left(\frac{p_1}{q_1}\right)^{l_1}\left(\frac{p_2}{q_2}\right)^{l_2} \cdots \left(\frac{p_m}{q_m}\right)^{l_m} \left(\frac{1-\sum_{i=1}^m q_i}{1-\sum_{i=1}^m p_i}\right)^{\sum_{i=1}^m l_i} \\
&&\qquad q_1^{k_1} q_2^{k_2} \cdots q_m^{k_m} \left(1-\sum_{i=1}^{m} p_i\right)^{2 {n/2 \choose 2}} \left(1-\sum_{i=1}^{m} q_i\right)^{\frac{n^2}{4}-\sum_{i=1}^m k_i} \\
&\stackrel{(a)}=& C(1+o(1)) \left(\frac{p_1}{q_1}\right)^{l_1}\left(\frac{p_2}{q_2}\right)^{l_2} \cdots \left(\frac{p_m}{q_m}\right)^{l_m},
\end{eqnarray*}
where (a) holds since $\left(\frac{1-\sum_{i=1}^m q_i}{1-\sum_{i=1}^m p_i}\right)^{\sum_{i=1}^m l_i}$ converges to $1$ as $n$ tends to $\infty$, and the remaining terms are fixed if the graph $G$ is given.
Therefore, ML rule finds two communities that maximizes $\left(\frac{p_1}{q_1}\right)^{l_1}\left(\frac{p_2}{q_2}\right)^{l_2} \cdots \left(\frac{p_m}{q_m}\right)^{l_m}$. In other words, it maximizes the weighted sum of the number of inner edges, $\sum_{i=1}^m l_i \log{\frac{p_i}{q_i}}$.

Using this result, we will get a sufficient and necessary condition of the event that ML estimator fails to detect the communities. For convenience, we define the following events for the true community assignment $A$ and $B$.
\begin{displaymath}
\left\{\begin{array}{ll}
F=\{\textrm{the maximum likelihood rule fails}\}\\
F_{A}=\{\sum_{i=1}^{m} E_i[v,A]\log{\frac{p_{i}}{q_{i}}}<\sum_{i=1}^{m} E_i[v,B]\log{\frac{p_{i}}{q_{i}}}, \exists v\in A\}\\
F_{B}=\{\sum_{i=1}^{m} E_i[v,A]\log{\frac{p_{i}}{q_{i}}}>\sum_{i=1}^{m} E_i[v,B]\log{\frac{p_{i}}{q_{i}}}, \exists v\in B\}\\
\end{array} \right.
\end{displaymath}
where $E_i[v,S]$ is the number of $i$-edges between the vertex $v$ and the set $S$.

First, let's show $F_A\cap F_B \Rightarrow F$. Suppose $F_A$ and $F_B$ happen simultaneously. Then, there exist vertices $v_A$ and $v_B$ such that,
\begin{eqnarray*}
v_A \in A &\text{and}& \sum_{i=1}^{m} E_i[v_A,A]\log{\frac{p_{i}}{q_{i}}}<\sum_{i=1}^{m} E_i[v_A,B]\log{\frac{p_{i}}{q_{i}}}\\
v_B \in B &\text{and}& \sum_{i=1}^{m} E_i[v_B,A]\log{\frac{p_{i}}{q_{i}}}>\sum_{i=1}^{m} E_i[v_B,B]\log{\frac{p_{i}}{q_{i}}}.
\end{eqnarray*}
Let's fix such $v_A$ and $v_B$. And we define a function $l_i (S)$ as the number of inner edges of type $i$ inside the set of nodes $S$. Also, we define a new community assignment, $\tilde{A}=(A\setminus v_A)\cup v_B$ and $\tilde{B}=(B\setminus v_B)\cup v_A$. Then, we can compare the weighted sum of the number of inner edges of $(A,B)$ and $(\tilde{A},\tilde{B})$:
\begin{eqnarray*}
\sum_{i=1}^m \left(l_i(\tilde{A})+l_i(\tilde{B})\right) \log{\frac{p_i}{q_i}} &=& \sum_{i=1}^m \left((l_i(A)+l_i(B)\right) \log{\frac{p_i}{q_i}} + \sum_{i=1}^m (E_i[v_B,A]-E_i[v_A,A]) \log{\frac{p_i}{q_i}}\\
&+& \sum_{i=1}^m (E_i[v_A,B]-E_i[v_B,B]) \log{\frac{p_i}{q_i}}\\
&>& \sum_{i=1}^m \left(l_i(A)+l_i(B)\right) \log{\frac{p_i}{q_i}}.
\end{eqnarray*}
Therefore, ML estimator does not choose the original community assignment $A$ and $B$, resulting in the event $F$.

On the other hand, let's assume $F$ happens. We define a function $o_i(S_1,S_2)$ as the number of edges of type $i$ across the set of nodes $S_1$ and $S_2$. Then, we can get an upper bound on probability of $F$ by the following lemma.

\begin{lem}
Suppose $F$ happens. Then, there exist $k$ and sets $A_w \subset A$ and $B_w \subset B$ with $|A_w|=|B_w|=k$ for $0\leq k \leq \frac{n}{4}$ such that
\begin{eqnarray*}
\sum_{i=1}^m \left[o_i(A_w, B\setminus B_w)+o_i(B_w,A\setminus A_w)\right]\log{\frac{p_i}{q_i}}\geq\sum_{i=1}^{m} \left[o_i(A_w, A\setminus A_w)+o_i(B_w,B\setminus B_w)\right]\log{\frac{p_i}{q_i}}
\end{eqnarray*}
\end{lem}

\begin{IEEEproof}
This lemma can be proved in a similar way as lemma 5 in \cite{Abbe:2016}. Assume $F$ happens. Then, there exist $\hat{A}$ and $\hat{B}$ such that
\begin{eqnarray*}
\sum_{i=1}^m \left(l_i(\hat{A})+l_i(\hat{B})\right) \log{\frac{p_i}{q_i}}\geq \sum_{i=1}^m \left(l_i({A})+l_i({B})\right) \log{\frac{p_i}{q_i}},
\end{eqnarray*}
where $|A\cap\hat{A}|,|B\cap\hat{B}|\geq\frac{n}{4}$, without loss of generality.

Then, we can spilt the number of inner edges inside each community as follows.
\begin{eqnarray*}
l_i(A)&=&l_i(A\cap\hat{A}) + l_i(A\cap\hat{B}) + o_i(A\cap\hat{A},A\cap\hat{B})\\
l_i(B)&=&l_i(B\cap\hat{A}) + l_i(B\cap\hat{B}) + o_i(B\cap\hat{A},B\cap\hat{B})\\
l_i(\hat{A})&=&l_i(A\cap\hat{A}) + l_i(B\cap\hat{A}) + o_i(A\cap\hat{A},B\cap\hat{A})\\
l_i(\hat{B})&=&l_i(A\cap\hat{B}) + l_i(B\cap\hat{B}) + o_i(A\cap\hat{B},B\cap\hat{B})
\end{eqnarray*}

Therefore, we get
\begin{eqnarray*}
\sum_{i=1}^{m} \left(o_i(A\cap\hat{A},B\cap\hat{A})+o_i(A\cap\hat{B},B\cap\hat{B})\right)\log{\frac{p_i}{q_i}} \geq \sum_{i=1}^{m} \left(o_i(A\cap\hat{A},A\cap\hat{B})+o_i(B\cap\hat{A},B\cap\hat{B})\right)\log{\frac{p_i}{q_i}}.
\end{eqnarray*}
Let $A_w = A\cap \hat{B}$ and $B_w = B\cap \hat{A}$. Then, the above inequality can be rewritten as,
\begin{eqnarray*}
\sum_{i=1}^m \left[o_i(A_w, B\setminus B_w)+o_i(B_w,A\setminus A_w)\right]\log{\frac{p_i}{q_i}}\geq\sum_{i=1}^{m} \left[o_i(A_w, A\setminus A_w)+o_i(B_w,B\setminus B_w)\right]\log{\frac{p_i}{q_i}}.
\end{eqnarray*}
\end{IEEEproof}
Now we define the probability $P_n^{(k)}$ as,
\begin{eqnarray*}
P_n^{(k)}=\Pr\left( \sum_{i=1}^m \left[o_i(A_w, B\setminus B_w)+o_i(B_w,A\setminus A_w)\right]\log{\frac{p_i}{q_i}}\geq\sum_{i=1}^{m} \left[o_i(A_w, A\setminus A_w)+o_i(B_w,B\setminus B_w)\right]\log{\frac{p_i}{q_i}} \right),
\end{eqnarray*}
where $k$ is size of $A_w$ and $B_w$.

Finally by applying union bound, we get an upper bound on $\Pr(F)$,
\begin{eqnarray*}
\Pr(F) \leq \sum_{k=1}^{n/4} \binom{n/2}{k}^2 P_n^{(k)}
\end{eqnarray*}

\section{Limit on the Community Detection}
In this section, we provide a fundamental limit of the community detection by proving two theorems. If $p_i = q_i$ for all $i$, the ML rule fails in detecting the communities obviously. Therefore, we focus on the case where there exists at least one $i$ such that $p_i \neq q_i$ throughout the paper.
	
\subsection{Achievability Proof}
\begin{thm}
If $\sum_{i=1}^{m} {(\sqrt{\alpha_{i}}-\sqrt{\beta_{i}})}^{2}>2$, then the maximum likelihood estimator detects the communities exactly with high probability.
\end{thm}
Note that $\sum_{i=1}^{m} {(\sqrt{\alpha_{i}}-\sqrt{\beta_{i}})}^{2}$ is related to the Hellinger distance between the two probability distributions $p$ and $q$ as
\begin{eqnarray*}
H^2 (p\|q)=\frac{\log n}{n}\sum_{i=1}^{m} {(\sqrt{\alpha_{i}}-\sqrt{\beta_{i}})}^{2}+o\left(\frac{\log n}{n}\right),
\end{eqnarray*}
where $H(p\|q)=\sqrt{\sum_{i=1}^{m+1} {(\sqrt{p_{i}}-\sqrt{q_{i}})}^{2}}$ is the Hellinger distance \cite{Kazakos:1980}. This distance is the special case of the CH-divergence given in \cite{Abbe:2015}.

	Before proving the theorem, we introduce some assumptions and definitions. For simplicity, we assume that $A=\{1,2,\dots,n/2\}$ and $B=\{n/2+1,\dots,n\}$. Also, we define independent random variables $W_{ij}$ for $1\leq i,j\leq n/2$ or $n/2< i,j\leq n$ such that
\begin{displaymath}
W_{ij} = \left\{ \begin{array}{ll}
\log{\frac{p_{k}}{q_{k}}} & \textrm{w.p. $p_{k}$ \qquad for $k=1,2,\dots ,m$}\\
0 & \textrm{w.p. $1-\sum_{k=1}^m p_{k}$}
\end{array} \right.
\end{displaymath}
and $Z_{ij}$ for $1\leq i\leq n/2$ and $n/2+1\leq j\leq n$ such that
\begin{displaymath}
Z_{ij} = \left\{ \begin{array}{ll}
\log{\frac{p_{k}}{q_{k}}} & \textrm{w.p. $q_{k}$ \qquad for $k=1,2,\dots,m$}\\
0 & \textrm{w.p. $1-\sum_{k=1}^m q_{k}$}
\end{array} \right.
\end{displaymath}

\begin{IEEEproof}[Proof of Theorem 1]
We can rewrite $P_n^{(k)}$ using $W_{ij}$ and $Z_{ij}$,
\begin{eqnarray}\label{eqn:upper}
P_n^{(k)} &=& \Pr\left( \sum_{i=1}^m \left[o_i(A_w, B\setminus B_w)+o_i(B_w,A\setminus A_w)\right]\log{\frac{p_i}{q_i}}\geq\sum_{i=1}^{m} \left[o_i(A_w, A\setminus A_w)+o_i(B_w,B\setminus B_w)\right]\log{\frac{p_i}{q_i}} \right)\nonumber\\
&=& \Pr\left( \sum_{i\in A_w, j \in B \setminus B_w}Z_{ij} + \sum_{i\in A\setminus A_w, j \in B_w}Z_{ij} \geq \sum_{i\in A_w, j \in A \setminus A_w}W_{ij} + \sum_{i\in B_w, j \in B\setminus B_w}W_{ij}\right)\nonumber\\
&=& \Pr \left( \sum_{i'=1}^{2k(\frac{n}{2}-k)} \left( Z_{i'}-W_{i'} \right) \geq 0 \right),
\end{eqnarray}
where $W_{i'}$ and $Z_{i'}$ are i.i.d. random variables that have the same distribution with $W_{ij}$ and $Z_{ij}$, respectively.

And we can get an upper bound on (\ref{eqn:upper}) by proving Lemma 1.

\begin{lem}
\begin{eqnarray*}
&&\Pr\left(\sum_{i'=1}^{2k(\frac{n}{2}-k)} \left( Z_{i'}-W_{i'} \right) \geq 0\right) \leq 2\exp\left[-2k\left(\frac{n}{2}-k\right)\left(\frac{\log{n}}{n}\sum_{i=1}^{m} {(\sqrt{\alpha_{i}}-\sqrt{\beta_{i}})}^{2}-C\frac{\log^2{n}}{n^2}-o\left(\frac{\log^2{n}}{n^2}\right)\right)\right].
\end{eqnarray*}
for some constant C.
\end{lem}
\begin{IEEEproof}
By the proof of Cramer's theorem in \cite{Dembo:2010}, for i.i.d. sequence $X_{i}$ and any closed set $F\subseteq {\rm I\!R}$,
$$\Pr\left(\frac{1}{n}\sum_{i=1}^{n}X_{i}\in F\right)\leq 2\exp\left(-n\inf_{x\in F}I(x)\right),$$
where $I(a)=\sup_{\theta\in{\rm I\!R}}(\theta a-\log{E[e^{\theta X}]})$. The proof for this theorem is in the appendix.\\

We need to evaluate the right-hand side of the following:
\begin{eqnarray}\label{eqn:cra1}
\Pr\left(\sum_{i'=1}^{2k(\frac{n}{2}-k)} \left( Z_{i'}-W_{i'} \right) \geq 0\right) \leq 2\exp\left(-2k\left(\frac{n}{2}-k\right)\inf_{x\geq 0}I(x)\right).
\end{eqnarray}

By direct computation, we can get the moment generating function of $(Z_{i'}-W_{i'})$, i.e.,
\begin{eqnarray*}
E\left[\exp\left(\theta\left(Z_{i'}-W_{i'}\right)\right)\right]=
\exp\left[-\frac{\log{n}}{n}\sum_{i=1}^{m}(\alpha_{i}+\beta_{i}-\alpha_{i}^{\theta}\beta_{i}^{1-\theta}-\beta_{i}^{\theta}\alpha_{i}^{1-\theta})+D(\theta)\frac{\log^2{n}}{n^2}+o\left(\frac{\log^2{n}}{n^2}\right)\right],
\end{eqnarray*}
where $D(\theta)$ is a function of $\theta$, the exact form of $D(\theta)$ is in Appendix 5.2.

Then, the right-hand side of (\ref{eqn:cra1}) is, for any $2\leq j\leq n/2$,
\begin{eqnarray}\label{eqn:upper2}
&-&\inf_{x\geq 0}\left[\sup_{\theta\in{\rm I\!R}}\left(\theta x-\log{E\left[\exp\left(\theta(Z_{i'}-W_{i'})\right)\right]}\right)\right]\nonumber\\
&=&-\inf_{x\geq 0}\Bigg[\sup_{\theta\in{\rm I\!R}}\left(\theta x +\frac{\log{n}}{n}\sum_{i=1}^{m}(\alpha_{i}+\beta_{i}-\alpha_{i}^{\theta}\beta_{i}^{1-\theta}-\beta_{i}^{\theta}\alpha_{i}^{1-\theta}) - D(\theta)\frac{\log^2{n}}{n^2}-o\left(\frac{\log^2{n}}{n^2}\right) \right)\Bigg]\nonumber\\
&\stackrel{(a)}\leq&-\inf_{x\geq 0}\Bigg[0.5x+\frac{\log{n}}{n}\sum_{i=1}^{m} {(\sqrt{\alpha_{i}}-\sqrt{\beta_{i}})}^{2} 
 -C\frac{\log^2{n}}{n^2}-o\left(\frac{\log^2{n}}{n^2}\right)\Bigg]\nonumber\\
&=&-\frac{\log{n}}{n}\sum_{i=1}^{m} {(\sqrt{\alpha_{i}}-\sqrt{\beta_{i}})}^{2}+C\frac{\log^2{n}}{n^2}+o\left(\frac{\log^2{n}}{n^2}\right),
\end{eqnarray}
where (a) holds by taking $\theta=0.5$ instead of taking the supremum and $D(\theta)=C$.

Finally, by combining (\ref{eqn:cra1}) and (\ref{eqn:upper2}), we get
\begin{eqnarray*}
\Pr\left(\sum_{i'=1}^{2k(\frac{n}{2}-k)} \left( Z_{i'}-W_{i'} \right) \geq 0\right) \leq 2\exp\left[-2k\left(\frac{n}{2}-k\right)\left(\frac{\log{n}}{n}\sum_{i=1}^{m} {(\sqrt{\alpha_{i}}-\sqrt{\beta_{i}})}^{2}-C\frac{\log^2{n}}{n^2}-o\left(\frac{\log^2{n}}{n^2}\right)\right)\right].
\end{eqnarray*}
\end{IEEEproof}

Now, the following is similar with proof of theorem 2 in \cite{Abbe:2016}. If we assume that $\sum_{i=1}^{m} {(\sqrt{\alpha_{i}}-\sqrt{\beta_{i}})}^{2}\geq 2+\epsilon$ for $\epsilon>0$,
\begin{eqnarray*}
\Pr(F)&\leq& \sum_{k=1}^{n/4} \binom{n/2}{k}^2 P_n^{(k)}\\
&\leq& 2\sum_{k=1}^{n/4} \binom{n/2}{k}^2 \exp\left[-2k\left(\frac{n}{2}-k\right)\left(\frac{\log{n}}{n}\sum_{i=1}^{m} {(\sqrt{\alpha_{i}}-\sqrt{\beta_{i}})}^{2}-C\frac{\log^2{n}}{n^2}-o\left(\frac{\log^2{n}}{n^2}\right)\right)\right]\\
&\stackrel{(a)}\leq& 2\sum_{k=1}^{n/4} \exp\left[2k\left(\log{\frac{n}{2k}}+1\right) - 2k\left(\frac{n}{2}-k\right)\left(\frac{\log{n}}{n}\sum_{i=1}^{m} {(\sqrt{\alpha_{i}}-\sqrt{\beta_{i}})}^{2}-C\frac{\log^2{n}}{n^2}-o\left(\frac{\log^2{n}}{n^2}\right)\right)\right]\\
&\stackrel{(b)}\leq& 2\sum_{k=1}^{n/4} \exp\left[2k\left( \log{n}-\log{2k}+1 -\left(\frac{1}{2}-\frac{k}{n}\right)\left((2+\epsilon)\log{n} -C\frac{\log^2{n}}{n}-o\left(\frac{\log^2{n}}{n}\right) \right) \right)\right]\\
&\leq& 2\sum_{k=1}^{n/4} \exp\left[2k\left( -\log{2k}+1 +\frac{2k}{n}\log{n} -\epsilon\left(\frac{1}{2}-\frac{k}{n}\right)\log{n} +\left(\frac{1}{2}-\frac{k}{n}\right)\left(C\frac{\log^2{n}}{n}+o\left(\frac{\log^2{n}}{n}\right)\right) \right)\right]\\
&\stackrel{(c)}\leq& 2\sum_{k=1}^{n/4} \exp\left[2k\left( -\log{2k}+1 +\frac{2k}{n}\log{n} -\frac{\epsilon}{4}\log{n} +o(1) \right)\right]\\
&=& 2\sum_{k=1}^{n/4} n^{-\frac{k}{2}\epsilon}\exp\left[2k\left( -\log{2k}+1 +\frac{2k}{n}\log{n} +o(1) \right)\right],
\end{eqnarray*}
where (a) holds by Stirling's formula, $\binom{n}{k}\leq (ne/k)^k$, (b) holds by the assumption, and (c) holds since $k\leq n/4$.

For sufficiently large $n$ and $1\leq k \leq \frac{n}{4}$, we have:
\begin{eqnarray*}
\frac{2}{3}\frac{1}{2k}\log{2k}-o\left(\frac{1}{2k}\right)\geq \frac{1}{n}\log{n}.
\end{eqnarray*}
In other words, $\log{2k} -\frac{2k}{n}\log{n} -o(1)\geq \frac{1}{3}\log{2k}$. By applying this inequality and $n^{-\frac{k}{2}\epsilon}\leq n^{-\frac{1}{2}\epsilon}$,
\begin{eqnarray*}
\Pr(F)\leq 2 n^{-\frac{1}{2}\epsilon}\sum_{k=1}^{n/4}\exp\left[ 2k\left(1-\frac{1}{3}\log{2k}\right) \right]
\end{eqnarray*}
Since $\sum_{k=1}^{n/4}\exp\left[ 2k\left(1-\frac{1}{3}\log{2k}\right) \right]=O(1)$, this proves Theorem 1.
\end{IEEEproof}

\subsection{Converse Proof}
\begin{thm}
If $\sum_{i=1}^{m} {(\sqrt{\alpha_{i}}-\sqrt{\beta_{i}})}^{2}<2$, then the probability that the maximum likelihood estimator fails to detect the communities does not go to zero.
\end{thm}
\begin{IEEEproof}[Proof of Theorem 2]
We will prove this theorem  via several lemmas. In this proof, we assume that there exists at least one $i$ such that $p_i>q_i$. Even if there is no such $i$, we can prove the theorem in a similar manner.
\begin{lem}
For $v\in H := \{1,\dots , n/\log^3 n\}$ and $M=\max_k \log{p_k/q_k}$, if
\begin{eqnarray*}
\Pr\left(\sum_{i=n/2+1}^{n}Z_{vi}-\sum_{i=\frac{n}{\log^3 n}+1}^{\frac{n}{2}}W_{vi}\geq\frac{M\log{n}}{\log{\log{n}}}\right) >n^{-1}\log^3 n\log{10},
\end{eqnarray*}
then, for sufficiently large $n$, $\Pr(F)\geq \frac{1}{3}$.
\end{lem}
\begin{IEEEproof}
This can be proved in a similar way as Lemma 3 in \cite{Abbe:2016}. Therefore, we describe its steps briefly.
We define the following events.
\begin{displaymath}
\left\{\begin{array}{ll}
\Delta=\{\forall v \in H,\sum_{i=1}^{m} E_i[v,H]\log{\frac{p_{i}}{q_{i}}}<\frac{M\log{n}}{\log{\log{n}}}\}\\
F_{H}^{(v)}=\{ v\in H \textrm{ satisfies } \sum_{i=1}^{m} E_i[v,A\setminus H]\log{\frac{p_{i}}{q_{i}}}+\frac{M\log{n}}{\log{\log{n}}} \leq \sum_{i=1}^{m} E_i[v,B]\log{\frac{p_{i}}{q_{i}}}\}\\
F_{H}=\{\cup_{v\in H} F_{H}^{(v)}\}\\
\end{array} \right.
\end{displaymath}
Then, the condition in the statement can be written as $\Pr(F_{H}^{(v)})>n^{-1}\log^3 n\log{10}$.

First, we show if $\Pr(F_{H}^{(v)})>n^{-1}\log^3 n\log{10}$, then $\Pr(F_H)\geq\frac{9}{10}$. Since $\Pr(F_H)=1-(1-\Pr(F_{H}^{(v)}))^{\frac{n}{\log^3 n}}$, we can easily check that $\Pr(F_H)\geq\frac{9}{10}$.

Also, we know that $(\Delta\cap F_{H})$ implies $F_{A}$. Therefore, to evaluate a lower bound on $\Pr(F_A)$, we should evaluate a lower bound on $\Pr(\Delta)$.
Define a new event $\Delta_{v}$ as $\Delta_{v}= \{\sum_{i=1}^{m} E_i[v,H]\log{\frac{p_{i}}{q_{i}}}<\frac{M\log{n}}{\log{\log{n}}}\}$, then
\begin{eqnarray*}
\Pr(\Delta_{v}^{c})=\Pr\left(\sum_{i=1}^{v-1} W_{iv}+\sum_{j=v+1}^{\frac{n}{\log^{3}{n}}} W_{vj} \geq \frac{M\log{n}}{\log{\log{n}}}\right).
\end{eqnarray*}

Now, we define new i.i.d. random variables $X_j\sim \operatorname{Bern}\left(\frac{\log{n}}{n}\sum_{i}\alpha_i\right)$. Then the following inequality holds:
\begin{eqnarray*}
\Pr\left(\sum_{i=1}^{v-1} W_{iv}+\sum_{j=v+1}^{\frac{n}{\log^{3}{n}}} W_{vj} \geq \frac{M\log{n}}{\log{\log{n}}}\right)
\leq \Pr\left(\sum_{j=1}^{\frac{n}{\log^{3}{n}}} X_{j} \geq \frac{\log{n}}{\log{\log{n}}}\right),
\end{eqnarray*}

By the multiplicative Chernoff bound,
\begin{eqnarray*}
\Pr\left(\sum_{j=1}^{\frac{n}{\log^{3}{n}}} X_{j} \geq \frac{\log{n}}{\log{\log{n}}}\right) \leq \left(\frac{\log^3{n}}{e\sum_{i}\alpha_{i}\log{\log{n}}}\right)^{-\frac{\log{n}}{\log{\log{n}}}}.
\end{eqnarray*}
By using the union bound, we get
\begin{eqnarray*}
1-\Pr(\Delta)&\leq&\frac{n}{\log^3 {n}}\Pr(\Delta_{i}^{c})\\
&\leq&\frac{n}{\log^3 n}\Pr\left(\sum_{j=1}^{\frac{n}{\log^3 n}} X_{j} \geq \frac{\log{n}}{\log{\log{n}}}\right)\\
&=&\exp\Bigg[\log{\frac{n}{\log^3 {n}}}-\frac{\log{n}}{\log{\log{n}}} \log{\frac{(\log{n})^3}{e\log{\log{n}}\sum_{i}\alpha_{i}}}\Bigg]\\
&=&\exp\left[-2\log{n}+o(\log n)\right].
\end{eqnarray*}•

Therefore, for sufficiently large $n$, $\Pr(\Delta)\geq\frac{9}{10}$. This implies $\Pr(F_{A})\geq \frac{2}{3}$, since $(\Delta\cap F_{H})$ implies $F_{A}$. Finally, since $(F_{A}\cap F_{B})$ implies $F$, we can conclude that if $\Pr(F_{A})\geq \frac{2}{3}$, then $\Pr(F)\geq \frac{1}{3}$.

\end{IEEEproof}

\begin{lem}
For sufficiently large $n$ and any positive constant $M$,
\begin{eqnarray*}
\Pr\left(\sum_{j=\frac{n}{\log^3n}+1}^{\frac{n}{2}}(Z_{1(j+\frac{n}{2})}-W_{1j})\geq \frac{M\log{n}}{\log{\log{n}}}+ o\left( \frac{\log{n}}{\log{\log{n}}}\right)\right)\\
\geq \exp\left[-\frac{\log{n}}{2}\left(\sum_{i=1}^{m} {\left(\sqrt{\alpha_{i}}-\sqrt{\beta_{i}}\right)}^{2}\right)-o\left(\log{n}\right)\right].
\end{eqnarray*}
\end{lem}
\begin{IEEEproof}
By the proof of Cramer's theorem in \cite{Dembo:2010}, for i.i.d. sequence $X_{i}$ and any open set $U\subseteq {\rm I\!R}$,
\begin{eqnarray}\label{eqn:lower}
\Pr\left(\frac{1}{n}\sum_{i=1}^{n}X_{i}\in U\right)
&\geq&\Pr\left(\frac{1}{n}\sum_{i=1}^{n}X_{i}\in (a-\epsilon,a+\epsilon)\right)\nonumber\\
&\geq&e^{-n(I(a)-\epsilon \mid \eta^* \mid)}\left(1-\frac{\sigma^2}{n\epsilon^2}\right),
\end{eqnarray}
where $a$ and $\epsilon$ are constants satisfing $(a-\epsilon,a+\epsilon)\subset U$, $\eta^*$ is the constant that minimizes $\log{E[e^{\eta X_i}]}$, $\sigma^2$ is the variance of $\tilde{X}_i$ and $I(a)=\sup_{\theta\in{\rm I\!R}}(\theta a-\log{E[e^{\theta X}]})$. The proof for this theorem is in the appendix.

	Here, $\tilde{X}_i$ is a random variable that follows $\frac{e^{\eta^* x}p(x)}{E[e^{\eta^* X}]}$ where $p(x)$ is the distribution of $X_i$. Since $\eta^*$ is a constant and $\sigma^2$ is of order of $\log n/n$, if we take $\epsilon=\log^{\frac{2}{3}} n/n$, $e^{n\epsilon\mid\eta^*\mid}$ and the last term in (\ref{eqn:lower}) is negligible.

Let $l=\frac{n}{2}-\frac{n}{\log^3n}$ for simplicity and take $a=\frac{M\log{n}}{l\log{\log{n}}} + \frac{1}{l}o\left(\frac{\log{n}}{\log{\log{n}}}\right) +\frac{\log^\frac{2}{3} n}{n}$. Then, we have,

\begin{eqnarray}\label{eqn:lower2}
\Pr\left(\sum_{j=\frac{n}{\log^3n}+1}^{\frac{n}{2}}(Z_{1(j+\frac{n}{2})}-W_{1j})\geq \frac{M\log{n}}{\log{\log{n}}}+ o\left(\frac{\log{n}}{\log{\log{n}}}\right)\right)\geq e^{-l(I(a)-\epsilon \mid \eta^* \mid)}\left(1-\frac{\sigma^2}{l\epsilon^2}\right).
\end{eqnarray}

Therefore, we need to evaluate $I(a)$ to evaluate the right-hand side of (\ref{eqn:lower2}).
\begin{eqnarray}
&&I(a)=\sup_{\theta\in{\rm I\!R}}\left(\theta a-\log{E[e^{\theta (Z_{1(j+n/2)}-W_{1j})}]}\right)\nonumber\\
&=&\sup_{\theta\in{\rm I\!R}}\Bigg(\theta a +\frac{\log{n}}{n}\sum_{i=1}^{m}(\alpha_{i}+\beta_{i}-\alpha_{i}^{\theta}\beta_{i}^{1-\theta}-\beta_{i}^{\theta}\alpha_{i}^{1-\theta}) -o\left(\frac{\log{n}}{n}\right)\Bigg)\nonumber\\
&\stackrel{(a)}=&\frac{\log{n}}{n}\sum_{i=1}^{m} {(\sqrt{\alpha_{i}}-\sqrt{\beta_{i}})}^{2} + o\left(\frac{\log{n}}{n}\right),\nonumber
\end{eqnarray}
where (a) holds since the function in the supremum is concave and the derivative of the function becomes zero when $\theta=0.5+\epsilon$ for $0\leq\epsilon<1/\log{\log\log{n}}$. Detailed explanation is in the Appendix.

\end{IEEEproof}

\begin{lem}
For some constant $K>0$,
\begin{eqnarray*}
\Pr\left(\sum_{i=1}^{\frac{n}{\log^3 n}} Z_{1(j+\frac{n}{2})} \geq \frac{1}{\log^2 n}\sum_{i=1}^{m}\beta_i \log\frac{\alpha_i}{\beta_i} -K\right)
\geq1-\frac{C}{K^2 \log^2 n}.
\end{eqnarray*}
\end{lem}

\begin{IEEEproof}
To prove this lemma, we will use Chebyshev's inequality. Since the variance of $Z_{1(j+\frac{n}{2})}$ is of order of $\log n/n$, $\sigma_{Z}^{2} \leq C\log n/n$ for some $C>0$. Then, we get,
\begin{eqnarray*}
&&\Pr\left(\sum_{i=1}^{\frac{n}{\log^3 n}} Z_{1(j+\frac{n}{2})}\geq \frac{1}{\log^2 n}\sum_{i=1}^{m}\beta_i \log\frac{\alpha_i}{\beta_i} -K\right)\\
&=&\Pr\left(\sum_{i=1}^{\frac{n}{\log^3 n}} Z_{1(j+\frac{n}{2})}\geq \frac{n}{\log^3 n}\left(E[Z_{1(1+\frac{n}{2})}]- \frac{\log^3 n}{n}K\right)\right)\\
&\geq& 1-\frac{C}{K^2 \log^2 n}.
\end{eqnarray*}
\end{IEEEproof}

Now, we assume that $\sum_{i=1}^{m} {(\sqrt{\alpha_{i}}-\sqrt{\beta_{i}})}^{2}\leq 2-\epsilon$ for $\epsilon>0$.
For simplicity, we define an event $S$ such that
\begin{eqnarray*}
S=\left\{\sum_{i=1}^{\frac{n}{\log^3 n}} Z_{1(j+\frac{n}{2})} \geq \frac{1}{\log^2 n}\sum_{i=1}^{m}\beta_i \log\frac{\alpha_i}{\beta_i} -K\right\}.
\end{eqnarray*}
Then, we have,
\begin{eqnarray}\label{eqn:result}
&&\Pr\left(\sum_{i=n/2+1}^{n}Z_{1i}-\sum_{i=\frac{n}{\log^3 n}+1}^{\frac{n}{2}}W_{1i}\geq\frac{M\log{n}}{\log{\log{n}}}\right)\nonumber\\
&\geq&\Pr\left(\sum_{i=n/2+1}^{n}Z_{1i}-\sum_{i=\frac{n}{\log^3 n}+1}^{\frac{n}{2}}W_{1i}\geq\frac{M\log{n}}{\log{\log{n}}}\mid S\right)\Pr(S)\nonumber\\
&\geq&\Pr\Bigg(\sum_{j=\frac{n}{\log^3n}+1}^{\frac{n}{2}}(Z_{1(j+\frac{n}{2})}-W_{1j})\geq \frac{M\log{n}}{\log{\log{n}}} -\frac{1}{\log^2 n}\sum_{i=1}^{m}\beta_i \log\frac{\alpha_i}{\beta_i} +K\Bigg)\Pr(S)\nonumber\\
&\stackrel{(a)}\geq& \exp\left[-\frac{\log{n}}{2}\left(\sum_{i=1}^{m} {\left(\sqrt{\alpha_{i}}-\sqrt{\beta_{i}}\right)}^{2}\right)-o\left(\log{n}\right)\right]\Pr(S)\nonumber\\
&\stackrel{(b)}\geq&n^{-1+\frac{\epsilon}{2}}\left( 1-\frac{C}{K^2 \log^2 n}\right)\nonumber\\
&\stackrel{(c)}>&n^{-1}\log^3 n\log{10},
\end{eqnarray}
where (a) holds by Lemma 3, (b) holds by the assumption and Lemma 4, and (c) holds for sufficiently large $n$.

Finally, by observing (\ref{eqn:result}), we know that $\Pr(F)\geq \frac{1}{3}$ by Lemma 2. This proves Theorem 2.

\end{IEEEproof}

\section{Appendix}
\subsection{Cramer's Theorem}
Cramer's theorem played a crucial role when we prove the main theorems of this paper. Therefore, in this section, we introduce how to prove this theorem \cite{Dembo:2010}. The following theorem is a slightly modified version of Cramer's theorem, but this is essentially the same as the original one.
\begin{thm*}[Cramer's Theorem]
For i.i.d. sequence $X_{i}$ and any closed set and open set $F,U\subseteq {\rm I\!R}$,
\begin{eqnarray}\label{eqn:app1}
\Pr\left(\frac{1}{n}\sum_{i=1}^{n}X_{i}\in F\right)\leq 2\exp\left(-n\inf_{x\in F}I(x)\right),
\end{eqnarray}
and
\begin{eqnarray}\label{eqn:app2}
\Pr\left(\frac{1}{n}\sum_{i=1}^{n}X_{i}\in U\right)
&\geq&\Pr\left(\frac{1}{n}\sum_{i=1}^{n}X_{i}\in (a-\epsilon,a+\epsilon)\right)\nonumber\\
&\geq&e^{-n(I(a)+\epsilon \mid \eta^* \mid)}\left(1-\frac{\sigma^2}{n\epsilon^2}\right),
\end{eqnarray}
where $I(a)=\sup_{\theta\in{\rm I\!R}}(\theta a-\log{E[e^{\theta X}]})$.
\end{thm*}
\begin{IEEEproof}
First, we will show (\ref{eqn:app1}). Let $F$ be a non-empty closed set. We can notice that (\ref{eqn:app1}) trivially holds when $\inf_{x\in F} I(x) = 0$. Hence, we assume $\inf_{x\in F} I(x) > 0$ through the following proof.
On the other hand, the following always  holds for all $x$ and $\theta \geq 0$,
\begin{eqnarray}\label{eqn:app3}
\Pr\left(\frac{1}{n}\sum_{i=1}^{n}X_{i}\geq x \right) &=& \operatorname{E}\left[\mathbbm{1}_{\frac{1}{n}\sum_{i=1}^{n}X_{i} -x \geq 0} \right] \leq \operatorname{E} \left[\exp\left(n\theta \left(\frac{1}{n}\sum_{i=1}^{n}X_{i} -x\right) \right) \right] \nonumber\\
&=& e^{-n\theta x} \prod_{i=1}^{n} \operatorname{E}\left[ e^{\theta X_i} \right] = \exp\left[-n\left(\theta x - \log\operatorname{E}\left[e^{\theta X}\right]\right)\right],
\end{eqnarray}
where the inequality holds by Chebycheff's inequality.

Now, we will show these two statements is true by proving Lemma \ref{eqn:app4}.
\begin{displaymath}
\left\{\begin{array}{ll}
\textrm{If } \bar{x}<\infty, \textrm{ for every } x>\bar{x}, \Pr\left(\frac{1}{n}\sum_{i=1}^{n}X_{i}\geq x\right)\leq e^{-nI(x)}\\
\textrm{If } \bar{x}>-\infty, \textrm{ for every } x<\bar{x}, \Pr\left(\frac{1}{n}\sum_{i=1}^{n}X_{i}\leq x\right)\leq e^{-nI(x)}\\
\end{array} \right.
\end{displaymath}

\begin{lem}\label{eqn:app4}
Assume $\bar{x}<\infty$. Then, for $x\geq \bar{x}$, $I(x)=\sup_{\theta\geq 0}\left[\theta x - \log \operatorname{E}\left[ e^{\theta X}\right]\right]$
\end{lem}
\begin{IEEEproof}
For all $\theta \in {\rm I\!R}$, we know that $\log\operatorname{E}\left[e^{\theta X}\right]\geq \operatorname{E}\left[\log e^{\theta X} \right]=\theta \bar{x}$ by Jensen's inequality.

If $\bar{x}=-\infty$, then $\log\operatorname{E}\left[e^{\theta X}\right]=\infty$ for negative $\theta$ so that this lemma trivially holds. Therefore, let's assume $\bar{x}$ is finite. Then, for $x\geq \bar{x}$ and $\theta<0$,
\begin{eqnarray*}
\theta x -\log\operatorname{E}\left[e^{\theta X}\right] \leq \theta\bar{x} - \log\operatorname{E}\left[e^{\theta X}\right] \leq 0
\end{eqnarray*}
This proves the lemma.
\end{IEEEproof}
By combining (\ref{eqn:app3}) and Lemma \ref{eqn:app4}, we can prove the first statement. Also, the second statement can be proved in a similar manner.

Now, since we are interested in the case of finite $\bar{x}$, we assume that $\bar{x}$ is finite. Then, $I(\bar{x}) = \sup_{\theta\geq 0}\left[\theta \bar{x} - \log \operatorname{E}\left[ e^{\theta \bar{x}}\right]\right]=0$ implies $\bar{x} \in F^c$ by the assumption. Let $(x_{-}, x_{+})$ be the union of all the open intervals $(a,b)\in F^c$ that contains $\bar{x}$. Since $F$ is non-empty, either $x_{-}$ or $x_{+}$ must be finite. If $x_{-}$ is finite, then $x_{-}\in F$, resulting in $\inf_{x\in F}I(x) \leq I(x_{-})$. Likewise, $\inf_{x\in F}I(x) \leq I(x_{+})$ whenever $x_{+}$ is finite. Finally,
\begin{eqnarray*}
\Pr\left(\frac{1}{n}\sum_{i=1}^{n}X_{i}\in F\right) &\leq& \Pr\left(\frac{1}{n}\sum_{i=1}^{n}X_{i}\leq x_{-} \right) + \Pr\left(\frac{1}{n}\sum_{i=1}^{n}X_{i}\geq x_{+}\right)\\
&\leq& e^{-nI(x_{-})} + e^{-nI(x_{+})} \leq 2 \exp\left(-n\inf_{x\in F}I(x)\right)
\end{eqnarray*}
This proves the first statement of the theorem.

To prove the second statement of the theorem, we need to show that for every $\epsilon >0$,
\begin{eqnarray*}
\Pr\left(\frac{1}{n}\sum_{i=1}^{n} Y_i \in (-\epsilon , \epsilon)\right)\geq \exp \left[-n\left(-\log \operatorname{E}\left[e^{\eta^* Y}\right] +\epsilon |\eta^*|\right) \right] \left(1- \frac{\sigma^2}{n \epsilon^2} \right),
\end{eqnarray*}
where $Y_i=X_i-a$.

Since if we rewrite the above inequality using $X_i$, we get:
\begin{eqnarray*}
\Pr\left(\frac{1}{n}\sum_{i=1}^{n} X_i \in (a-\epsilon , a+\epsilon)\right)&\geq& \exp \left[-n\left(-\log \operatorname{E}\left[e^{\eta^* (X-a)}\right] +\epsilon |\eta^*|\right) \right] \left(1- \frac{\sigma^2}{n \epsilon^2} \right)\\
&=& \exp \left[-n\left(\eta^* a-\log \operatorname{E}\left[e^{\eta^* X}\right] +\epsilon |\eta^*|\right) \right] \left(1- \frac{\sigma^2}{n \epsilon^2} \right)\\
&\geq& \exp \left[-n\left(I(a)+\epsilon |\eta^*|\right) \right] \left(1- \frac{\sigma^2}{n \epsilon^2} \right).
\end{eqnarray*}

Now, as $\log\operatorname{E}\left[e^{\lambda X}\right]$ is a continuous and differentiable function, there exists a finite $\eta$ such that
\begin{eqnarray*}
\log\operatorname{E}\left[e^{\eta X}\right] = \inf_{\lambda \in {\rm I\!R}} \log\operatorname{E}\left[e^{\lambda X}\right] \textrm{ and } \frac{d}{d\lambda} \log\operatorname{E}\left[e^{\lambda X}\right]\Big|_{\lambda=\eta}=0.
\end{eqnarray*}

Using $\eta$, we define new random variables, $\tilde{X_i}\sim\frac{e^{\eta x}p(x)}{\operatorname{E}\left[e^{\eta X}\right]}$. Then, the expected value of $\tilde{X}$ is,
\begin{eqnarray*}
\operatorname{E}[\tilde{X}]&=&\frac{1}{\operatorname{E}\left[e^{\eta X}\right]}\sum_{\chi} x_i e^{\eta x_i} p(x_i)\\
&=&\frac{1}{\operatorname{E}\left[e^{\eta X}\right]} \frac{d}{d\lambda} \log\operatorname{E}\left[e^{\lambda X}\right]\Big|_{\lambda=\eta}=0.
\end{eqnarray*}
Therefore, by the law of large numbers, $\Pr\left(\frac{1}{n} \sum_{i=1}^n \tilde{X_i}\in (-\epsilon, \epsilon)\right)\geq\left(1- \frac{\sigma^2}{n \epsilon^2} \right)$.

Finally, using the above results,
\begin{eqnarray*}
\Pr\left(\frac{1}{n} \sum_{i=1}^n {X_i}\in (-\epsilon, \epsilon)\right) &=& \sum_{|\sum x_i|<n\epsilon}p(x_1)p(x_2)\cdots p(x_n)\\
&\geq& e^{-n\epsilon |\eta|} \sum_{|\sum x_i|<n\epsilon}\exp\left(\eta \sum_{i=1}^n x_i\right)p(x_1)p(x_2)\cdots p(x_n)\\
&=& e^{-n\epsilon |\eta|} \exp\left(n\log\operatorname{E}\left[e^{\eta X}\right]\right) \Pr\left(\frac{1}{n} \sum_{i=1}^n \tilde{X_i}\in (-\epsilon, \epsilon)\right)\\
&\geq&\exp \left[-n\left(-\log \operatorname{E}\left[e^{\eta X}\right] +\epsilon |\eta|\right) \right] \left(1- \frac{\sigma^2}{n \epsilon^2} \right).
\end{eqnarray*}
This proves the second statement of the theorem.
\end{IEEEproof}

\subsection{Detailed Explanation for the proof of Lemma 3}
First, we should calculate the moment generating function of $(Z_{1(j+n/2)}-W_{1j})$. For simplicity, $\sum_{i=1}^{m}p_{i}$ and $\sum_{i=1}^{m}q_{i}$ are denoted as $p^{*}$ and $q^{*}$ respectively. $\alpha^{*}$ and $\beta^{*}$ are defined in a similar way.\\
Then we get

\begin{displaymath}
Z_{1(j+n/2)}-W_{1j} = \left\{ \begin{array}{ll}
0 & \textrm{w.p. $\sum_{k=1}^{m}p_{k}q_{k}+(1-p^{*})(1-q^{*})$}\\
\log{\frac{p_{i}}{q_{i}}} & \textrm{w.p. $(1-p^{*})q_{i}$}\\
\log{\frac{q_{i}}{p_{i}}} & \textrm{w.p. $p_{i}(1-q^{*})$}\\
\log{\frac{p_{1}q_{i}}{q_{1}p_{i}}} & \textrm{w.p. $p_{i}q_{1}$ \quad for $i\neq 1$}\\
\log{\frac{p_{2}q_{i}}{q_{2}p_{i}}} & \textrm{w.p. $p_{i}q_{2}$ \quad for $i\neq 2$}\\
\vdots\\
\log{\frac{p_{m}q_{i}}{q_{m}p_{i}}} & \textrm{w.p. $p_{i}q_{m}$ \quad for $i\neq m$}\\
\end{array} \right.
\end{displaymath}
and
\begin{eqnarray*}
E\big[e^{\theta (Z_{1(j+n/2)}-W_{1j})}\big]&=&\bigg[\sum_{i=1}^{m}p_{i}q_{i}+(1-p^{*})(1-q^{*})\bigg] + \sum_{i=1}^{m}(1-p^{*})q_{i}\left(\frac{p_{i}}{q_{i}}\right)^{\theta} \\ &&\qquad\qquad + \sum_{i=1}^{m}p_{i}(1-q^{*})\left(\frac{q_{i}}{p_{i}}\right)^{\theta} + \sum_{i\neq j}p_{i}q_{j}\left(\frac{p_{j}q_{i}}{q_{j}p_{i}}\right)^{\theta}\\
&=&\left(\frac{\log{n}}{n}\right)^{2}\sum_{i=1}^{m}\alpha_{i}\beta_{i} + \left(1-\frac{\alpha^{*}\log{n}}{n}\right)\left(1-\frac{\beta^{*}\log{n}}{n}\right) +  \sum_{i=1}^{m}\left(1-\frac{\alpha^{*}\log{n}}{n}\right)\frac{\beta_{i}\log{n}}{n}\left(\frac{\alpha_{i}}{\beta_{i}}\right)^{\theta}\\
&+& \sum_{i=1}^{m}\frac{\alpha_{i}\log{n}}{n}\left(1-\frac{\beta^{*}\log{n}}{n}\right)\left(\frac{\beta_{i}}{\alpha_{i}}\right)^{\theta} + \left(\frac{\log{n}}{n}\right)^{2}\sum_{i\neq j}\alpha_{i}\beta_{j}\left(\frac{\alpha_{j}\beta_{i}}{\beta_{j}\alpha_{i}}\right)^{\theta}\\
&=& 1-\frac{\alpha^{*}\log{n}}{n}-\frac{\beta^{*}\log{n}}{n}+\sum_{i=1}^{m}\bigg[\frac{\alpha_{i}\log{n}}{n}\left(\frac{\beta_{i}}{\alpha_{i}}\right)^{\theta}+\frac{\beta_{i}\log{n}}{n}\left(\frac{\alpha_{i}}{\beta_{i}}\right)^{\theta}\bigg]\\
&+&\left(\frac{\log{n}}{n}\right)^2 \bigg[\sum_{i=1}^{m}\alpha_i \beta_i + \alpha^{*} \beta^{*} - \sum_{i=1}^{m} \left(\alpha^{*} \beta_i \left(\frac{\alpha_i}{\beta_i}\right)^{\theta}+\beta^{*}\alpha_i \left( \frac{\beta_i}{\alpha_i}\right)^{\theta}\right) + \sum_{i\neq j}\alpha_{i}\beta_{j}\left(\frac{\alpha_{j}\beta_{i}}{\beta_{j}\alpha_{i}}\right)^{\theta}\bigg] \\
&=& 1 +C(\theta)\frac{\log n}{n} + D(\theta)\left(\frac{\log n}{n}\right)^2,
\end{eqnarray*}
where $C(\theta)=-\alpha^* -\beta^* +\sum_{i=1}^{m} \bigg[\alpha_i \left(\frac{\beta_i}{\alpha_i}\right)^{\theta} +\beta_i \left(\frac{\alpha_i}{\beta_i}\right)^{\theta}\bigg]$

and $D(\theta)=\sum_{i=1}^{m}\alpha_i \beta_i + \alpha^{*} \beta^{*} - \sum_{i=1}^{m} \left(\alpha^{*} \beta_i \left(\frac{\alpha_i}{\beta_i}\right)^{\theta}+\beta^{*}\alpha_i \left( \frac{\beta_i}{\alpha_i}\right)^{\theta}\right) + \sum_{i\neq j}\alpha_{i}\beta_{j}\left(\frac{\alpha_{j}\beta_{i}}{\beta_{j}\alpha_{i}}\right)^{\theta}$.

\begin{eqnarray}
I(a)&=&\sup_{\theta\in{\rm I\!R}}\left(\theta a-\log{E[e^{\theta (Z_{1(j+n/2)}-W_{1j})}]}\right)\nonumber\\
&=&\sup_{\theta\in{\rm I\!R}}\left(\theta a - \log{\left( 1+C(\theta)\frac{\log n}{n} + D(\theta) \left(\frac{\log n}{n}\right)^2 \right)}\right)\nonumber\\
&\stackrel{(a)}=&\frac{\log{n}}{n}\sum_{i=1}^{m} {(\sqrt{\alpha_{i}}-\sqrt{\beta_{i}})}^{2} + o\left(\frac{\log{n}}{n}\right),\nonumber
\end{eqnarray}

Define a function $f:{\rm I\!R}\mapsto{\rm I\!R}$ such that $$f(\theta)=\theta a - \log{\left( 1+C(\theta)\frac{\log n}{n} + D(\theta) \left(\frac{\log n}{n}\right)^2 \right)}.$$
Then the derivative and the second derivative of $f$ are given by
$$f'(\theta)=a - \frac{C'(\theta)\frac{\log n}{n} + D'(\theta) \left(\frac{\log n}{n}\right)^2}{1+C(\theta)\frac{\log n}{n} + D(\theta) \left(\frac{\log n}{n}\right)^2},$$
where $C'(\theta) = \sum_{i=1}^{m} \bigg[\alpha_i \left(\frac{\beta_i}{\alpha_i}\right)^{\theta}\log{\frac{\beta_i}{\alpha_i}} +\beta_i \left(\frac{\alpha_i}{\beta_i}\right)^{\theta} \log{\frac{\alpha_i}{\beta_i}} \bigg]$

and $D'(\theta) =  - \sum_{i=1}^{m} \left(\alpha^{*} \beta_i \left(\frac{\alpha_i}{\beta_i}\right)^{\theta} - \beta^{*}\alpha_i \left( \frac{\beta_i}{\alpha_i}\right)^{\theta}\right)\log{\frac{\alpha_i}{\beta_i}} + \sum_{i\neq j}\alpha_{i}\beta_{j}\left(\frac{\alpha_{j}\beta_{i}}{\beta_{j}\alpha_{i}}\right)^{\theta}\log{\frac{\alpha_{j}\beta_{i}}{\beta_{j}\alpha_{i}}}$

Also,
\begin{eqnarray*}
f''(\theta) &=& - \frac{\bigg[ 1+C(\theta)\frac{\log n}{n} + D(\theta) \left(\frac{\log n}{n}\right)^2 \bigg] \bigg[ C''(\theta)\frac{\log n}{n} + D''(\theta) \left(\frac{\log n}{n}\right)^2 \bigg] - \bigg[ C'(\theta)\frac{\log n}{n} + D'(\theta) \left(\frac{\log n}{n}\right)^2 \bigg]^2}{\bigg[ 1+C(\theta)\frac{\log n}{n} + D(\theta) \left(\frac{\log n}{n}\right)^2 \bigg]^2}\\
&=& - \frac{\log n}{n} \frac{\bigg[ 1+C(\theta)\frac{\log n}{n} + D(\theta) \left(\frac{\log n}{n}\right)^2 \bigg] \bigg[ C''(\theta) + D''(\theta) \frac{\log n}{n} \bigg] - \frac{\log n}{n} \bigg[ C'(\theta) + D'(\theta) \frac{\log n}{n} \bigg]^2}{\bigg[ 1+C(\theta)\frac{\log n}{n} + D(\theta) \left(\frac{\log n}{n}\right)^2 \bigg]^2},
\end{eqnarray*}
where $C''(\theta) = \sum_{i=1}^{m} \bigg[\alpha_i \left(\frac{\beta_i}{\alpha_i}\right)^{\theta} \left(\log{\frac{\beta_i}{\alpha_i}}\right)^2 +\beta_i \left(\frac{\alpha_i}{\beta_i}\right)^{\theta} \left(\log{\frac{\alpha_i}{\beta_i}}\right)^2 \bigg]$.

We can notice that there exists $\delta$ such that $C''(\theta)>\delta>0$ for all $\theta \in {\rm I\!R}$. Therefore, there exists $N_1$ such that for all $n>N_1$, $f''(\theta)<0$ for all $\theta \in {\rm I\!R}$.

Since $a= \frac{M\log{n}}{l\log{\log{n}}} + \frac{1}{l}o\left(\frac{\log{n}}{\log{\log{n}}}\right) +\frac{\log^\frac{2}{3} n}{n}$ and $C'(0.5)=0$,
$$f'(0.5)= \frac{M\log{n}}{l\log{\log{n}}} + \frac{1}{l}o\left(\frac{\log{n}}{\log{\log{n}}}\right) +\frac{\log^\frac{2}{3} n}{n} - \frac{D'(0.5) \left(\frac{\log n}{n}\right)^2}{1+C(0.5)\frac{\log n}{n} + D(0.5) \left(\frac{\log n}{n}\right)^2}.$$
Therefore, there exists $N_2$ such that for all $n>N_2$, $f'(0.5)>0$.

For $\theta = 0.5+1/\log\log\log{n}$,
\begin{eqnarray*}
C'(\theta) &=& \sum_{i=1}^{m} \bigg[\alpha_i \left(\frac{\beta_i}{\alpha_i}\right)^{0.5+\frac{1}{\log\log\log{n}}}\log{\frac{\beta_i}{\alpha_i}} +\beta_i \left(\frac{\alpha_i}{\beta_i}\right)^{0.5+\frac{1}{\log\log\log{n}}} \log{\frac{\alpha_i}{\beta_i}} \bigg]\\
&=& \sum_{i=1}^{m} \bigg[ \left(\frac{\beta_i}{\alpha_i}\right)^{\frac{1}{\log\log\log{n}}} - \left(\frac{\alpha_i}{\beta_i}\right)^{\frac{1}{\log\log\log{n}}} \bigg] \sqrt{\alpha_i \beta_i} \log{\frac{\alpha_i}{\beta_i}}\\
&=& \sum_{i=1}^{m} \bigg[ \exp \left( \frac{1}{\log\log\log{n}} \log{\frac{\alpha_i}{\beta_i}} \right) - \exp \left( -\frac{1}{\log\log\log{n}} \log{\frac{\alpha_i}{\beta_i}} \right) \bigg] \sqrt{\alpha_i \beta_i} \log{\frac{\alpha_i}{\beta_i}}\\
&\stackrel{(a)}=& \sum_{i=1}^{m} 2\bigg[ \frac{1}{\log\log\log{n}} \log{\frac{\alpha_i}{\beta_i}} + \left( \frac{1}{\log\log\log{n}} \log{\frac{\alpha_i}{\beta_i}} \right)^3 + \cdots \bigg] \sqrt{\alpha_i \beta_i} \log{\frac{\alpha_i}{\beta_i}}\\
&=& \frac{1}{\log\log\log{n}} \sum_{i=1}^{m} 2\bigg[ \left( \log{\frac{\alpha_i}{\beta_i}} \right)^2 + \left( \frac{1}{\log\log\log{n}} \right)^2 \left( \log{\frac{\alpha_i}{\beta_i}} \right)^4 + \cdots \bigg] \sqrt{\alpha_i \beta_i},
\end{eqnarray*}
where (a) holds by $e^x - e^{-x} = 2 \left( x + \frac{1}{3}x^3 + \cdots \right)$.

Hence, $f'(0.5+ 1/\log\log\log{n})$ is given by,
$$f'\left(0.5+\frac{1}{\log\log\log n}\right)=a - \frac{C'\left(0.5+\frac{1}{\log\log\log n}\right) \frac{\log n}{n} + D'\left(0.5+\frac{1}{\log\log\log n}\right) \left(\frac{\log n}{n}\right)^2}{1+C\left(0.5+\frac{1}{\log\log\log n}\right) \frac{\log n}{n} + D\left(0.5+\frac{1}{\log\log\log n}\right) \left(\frac{\log n}{n}\right)^2},$$

Since $a \ll \frac{\log{n}}{n\log\log\log{n}}$  and $ D'\left(0.5+\frac{1}{\log\log\log n}\right) \left(\frac{\log n}{n}\right)^2 \ll \frac{\log{n}}{n\log\log\log{n}}$  for sufficiently large $n$, there exists $N_3$ such that $f'(0.5+ 1/\log\log\log{n})<0$ for all $n>N_3$.

Fix $n >\max (N_1,N_2,N_3)$. Then, $f''(\theta)<0$ for all $\theta \in {\rm I\!R}$, $f'(0.5)>0$ and $f'(0.5+ 1/\log\log\log{n})<0$. By the continuity of $f'(\theta)$, we can conclude that $f'(\theta^{*})=0$ for $0.5\leq\theta^{*}\leq0.5+1/\log\log\log{n}$.

Let $\theta^{*}=0.5+\epsilon$ where $0\leq\epsilon\leq 1/\log\log\log{n}$.

\begin{eqnarray*}
f(\theta^{*})&=& (0.5+\epsilon)a - \log{\bigg[ 1 + C(0.5+\epsilon)\frac{\log n}{n} + D(0.5+\epsilon) \left(\frac{\log n}{n}\right)^2 \bigg]}\\
&\stackrel{(a)}=&(0.5+\epsilon)a - \bigg[ C(0.5+\epsilon)\frac{\log n}{n} + D(0.5+\epsilon) \left(\frac{\log n}{n}\right)^2 \bigg]\\ & &\qquad + \frac{1}{2(1+\xi)^2}\bigg[ C(0.5+\epsilon)\frac{\log n}{n} + D(0.5+\epsilon) \left(\frac{\log n}{n}\right)^2 \bigg]^2\\
&=& -C(0.5+\epsilon)\frac{\log n}{n} + o\left( \frac{\log{n}}{n} \right)\\
&=& \frac{\log{n}}{n} \sum_{i=1}^{m} \bigg[\alpha_{i}+\beta_{i}-\left(\left(\frac{\alpha_{i}}{\beta_{i}}\right)^{\epsilon}+\left(\frac{\beta_{i}}{\alpha_{i}}\right)^{\epsilon}\right)\sqrt{\alpha_{i}\beta_{i}}\bigg] + o\left(\frac{\log{n}}{n}\right)\\
&=& \frac{\log{n}}{n} \sum_{i=1}^{m}[\alpha_i + \beta_i - 2\sqrt{\alpha_i \beta_i}] + \frac{\log{n}}{n} \sum_{i=1}^{m}\left(2- \left(\frac{\alpha_i}{\beta_i}\right)^\epsilon -\left(\frac{\beta_i}{\alpha_i}\right)^\epsilon \right)\sqrt{\alpha_i \beta_i} + o\left( \frac{\log{n}}{n} \right)\\
&=& \frac{\log{n}}{n} \sum_{i=1}^{m} (\sqrt{\alpha_i}-\sqrt{\beta_i})^{2}+o\left( \frac{\log{n}}{n} \right),
\end{eqnarray*}
where (a) holds by Taylor's theorem, $\log{(1+x)}=x - \frac{1}{2(1+\xi)^2} x^2$, $\xi$ is between $0$ and $x$.

Since $n$ is arbitrary, (a) of the paper holds for sufficiently large $n$.


\section*{Acknowledgment}
This work was supported in part by the CISS funded by the Ministry of Science, ICT \& Future Planning as the Global Frontier Project.



%

\end{document}